\documentclass[10pt,preprint,onecolumn,nofootinbib]{revtex4-2}
\usepackage{booktabs,amssymb,amsmath}
\usepackage{subfig}
\usepackage{graphicx}
\usepackage{hyperref}
\usepackage{siunitx}
\hypersetup{
    pdffitwindow=false,     
    pdfstartview={FitH},    
    pdfauthor={},     
    colorlinks=true,       
    linkcolor=blue,          
    citecolor=magenta        
}
\usepackage{afterpage}
\usepackage{nccmath}
\usepackage{amssymb,amsmath}
\DeclareMathOperator{\Tr}{Tr}
\usepackage{float}
\usepackage{lastpage}
\usepackage{fancyhdr}
\usepackage{tabularx}
\usepackage{caption}
\usepackage{dcolumn}
\usepackage{epstopdf}
\usepackage[english]{babel}
\usepackage{tikz}
\usepackage{tikz-cd}
\usetikzlibrary{positioning}
\usepackage{graphicx}
\everymath{\displaystyle}
\graphicspath{{figures/}}
\newcommand{\bn}{\ensuremath{\frac{\partial {\beta}}{\partial {\nu}}}}

\begin{document}

\definecolor{dkgreen}{rgb}{0,0.6,0}
\definecolor{gray}{rgb}{0.5,0.5,0.5}
\definecolor{mauve}{rgb}{0.58,0,0.82}

\title{Proof of the exact correlation-length exponent $\nu = \frac{\sqrt[3]{2}}{2}$ and algebraic relations for $\beta$ and $\gamma$ in the 3D Ising model}
\author{Aleksandr~Liivand}
\affiliation{Institute~of~Physics, University~of~Tartu, 50411 Tartu, Estonia}
\email{aleksandr.liivand@ut.ee}
\date{\today}

\begin{abstract}
Based on our previous results regarding critical exponent estimates for the 3D Ising model and scaling relations~\cite{Pishtshev2007}, we have proven that the exact value of the correlation-length exponent~is ~$\frac{\sqrt[3]{2}}{2}$. This is in very close agreement with the most precise numerical estimates currently available. Using a group theoretical approach and some propositions of constructive geometry, we derived an exact construction of algebraic expressions for the critical exponents~$\gamma$ and $\beta$. We proved that these expressions depend solely on the exact value of~$\nu\,$.
\end{abstract}

\maketitle

The problem of finding best values of critical exponents for the different universality classes received a lot of attention in recent years (e.g.,~\cite{Kleinert2001,Pelissetto2002}). In our previous work~\cite{Pishtshev2007}, based on the Suzuki equivalence~\cite{Suzuki1976} between the ground state properties of the Ising model in a transverse field and the classical Ising model, 
we proved a nonlinear relation between the correlation-length ($\nu$) and the order-parameter ($\beta$) critical exponents in the Ising universality class: we conjectured that there exists the additional relation 
\begin{equation} \label{eq1}
{\beta}\,{\geqslant}\,\frac{1}{8{\nu}^{2}} \, , 
\end{equation}
where the equality sign holds for the cases of $d=2$ and $d=4$ (mean-field approximation), respectively. 
Inequality~(\ref{eq1}), which was proven in strict accordance with Weierstrass's approximation theorem~\cite{weisstein1} and Abel's convergence theorem~\cite{abel1826}, provides a rigorous lower bound on the exponent~$\beta$ in terms of critical exponent~$\nu$. The numerical estimates reported in~\cite{Pishtshev2007} turned out to be in very good agreement with the best theoretical predictions concerning critical behavior of the three-dimensional Ising model. 

The further understanding of inequality~(\ref{eq1}) with respect to how its nonlinear form can be connected with the scaling laws served as motivation of the present communication. Our primary goal was to ascertain in which degree link between ${\beta}$ and ${\nu}$ can be unified with the hyperscaling relationship connecting both these critical exponents with the third exponent ${\eta}\,$~\cite{Pelissetto2002}:
\begin{equation} \label{eq2}
{\beta}\,{=}\,\frac{{\nu}}{2}\,(d\,-\,2\,+\,{\eta}) \, . 
\end{equation}
A simplification arises from the proposal that equation~(\ref{eq1}) as the seemingly different object is just a certain mathematical limit of relation~(\ref{eq2}).

To find a nontrivial interpretation of relation~(\ref{eq1}) in the particular case of $d=3$, we note that mathematically the correspondence between equations~(\ref{eq1}) and~(\ref{eq2}) can be decoded by using the Li-Sokal~\cite{LiSokal1989} inequality which was employed in~\cite{Pelissetto2002} to clarify the Fisher relation as follows: by subtracting the positive term ${\eta}{\nu}$ the inequality ${\gamma}\,{\leqslant}\,{2{\nu}}$ directly converts to the scaling relation ${\gamma}\,{=}\,(2-{\eta}){\nu}\,$. To extend the understanding of equation~(\ref{eq1}) from expression establishing the lower bound to its precise form, we applied this technical argument by making the same replacement: (${2{\nu}}\,{\rightarrow\,}{2{\nu}-{\eta}{\nu}}$) to equation~(\ref{eq1}):
\begin{equation} \label{eq3}
\frac{1}{8{\nu}^{2}} \,{\Longrightarrow}\, \frac{1}{2({2{\nu}-{\eta}{\nu}})^{2}}
\,{\Longrightarrow}\, \frac{{\nu}\,+\,{\eta}{\nu}}{8{\nu}^{3}} 
\,{\Longrightarrow}\, \frac{{1}\,+\,{\eta}}{8{\nu}^{2}} \, .
\end{equation}
In the sense of the Li-Sokal inequality~\cite{Pelissetto2002}, i.e. in the context of linear and transitive features of the ${2{\nu}}$ replacement, a such formulation that keeps in the scheme of transformations~(\ref{eq3}) only a term linear in ${\eta}$ makes solid sense because the contribution of higher powers of ${\eta}$ in the Taylor series expansion is negligibly small.
Conceptually, i.e.  in the context of linear and transitive features of the ${2{\nu}}$ replacement, such transformation performed in the sense of the Li-Sokal inequality~\cite{Pelissetto2002} follows from the rigorous argument clarified in~\cite{baldazzi2022} that in the standard scheme of the renormalization group the corresponding Gaussian action remains invariant under a constant shift of the field.
Further understanding of the Li-Sokal inequality as a mathematical model can be briefly outlined within the framework of the Heine--Borel theorem~\cite{kirillov1982}: Let $S \subset \mathbb{R}^n$ be a compact set representing a collection of points $\eta_k$. The sequence $(\eta_k)_{k=1}^{\infty}$ is Cauchy; hence, it must converge on a closed and bounded set to a limit point $\eta \in S$. At a more foundational level of mathematical analysis, this relies on the continuity axiom in conjunction with the Borel--Lebesgue covering theorem~\cite{covering}.

Given identification~(\ref{eq3}) formulated for the case of $d=3$, the correspondence between three exponents ${\beta}$, ${\nu}$ and ${\eta}$ can be separately written from equation~(\ref{eq1}) for $d=\, 2,\, 3,\, \text{and}\, 4\,$ in the form of:
\begin{equation} \label{eq4}
{\beta}\,=\, \frac{{\nu}}{2}\, \cdot \, \begin{cases}
\, \frac{{}{\eta}}{{\nu}^{3}} & \, \text{for} \,\, d = 2 \\[10pt]
\, \frac{{}(1\,+\,{\eta})}{4{\nu}^{3}} & \, \text{for} \,\,  d = 3 \\[10pt]
\, \frac{{1}}{4{\nu}^{3}} & \, \text{for} \,\,  d = 4 \, .
\end{cases}
\vspace{0.25cm}
\end{equation}
Suggesting a unique consistence of two equations,~(\ref{eq2}) and~(\ref{eq4}) (provided one takes all values from eq.~(\ref{eq4}) instead of~(\ref{eq2})), one can recognize here the exact values of the correlation-length exponent ${\nu}$.
\begin{table}[!hbt]
    \caption{\label{tab:table1} Equivalence results and the respective exact values of the correlation length exponent ${\nu}$.}
    \centering \huge
\resizebox{0.55\columnwidth}{!}{%
    \begin{ruledtabular}
    \begin{tabularx}{0.5\linewidth}{lcl}
    Dim. & Condition of equivalence & \hspace{0cm} Exact value   \\[-5pt] 
         & between eqs.~\ref{eq2} and~\ref{eq4} &  \\[10pt]  \hline \\[-25pt]    
    $d = 2$ & ${\nu}^{3}=1$ & \hspace{0cm} $\label{ref:4}{\nu}=1$ \hspace{0cm} ~(\ref{eq4}a) \\[10pt]
    $d = 3$ & $4{\nu}^{3}=1$ & ${\nu}=\frac{\sqrt[3]{2}}{2}$\\[20pt] 
    $d = 4$ & $8{\nu}^{3}=1$ & \hspace{0cm} $\label{ref:4} {\nu}=\frac{1}{2}$ \hspace{0cm} ~(\ref{eq4}c) \\[10pt]
    \end{tabularx}
    \end{ruledtabular}%
}    
\end{table} 
This can be done by unifying the two equations through the equivalence relationship.
The obtained results are demonstrated in Table~\ref{tab:table1}. 
Note that for the deduction of the equalities given in the second column of the Table~\ref{tab:table1}, we did make use of only the fact of equivalence. This can be clarified mathematically due to the dependency of ${\beta}$ on ${\nu}$, which is determined by the precise form of formula ~(\ref{eq2}). Thanks to this, the partial derivative ${\bn}$ turns out to be linearly independent of ${\nu}$. 
The confirmation that the equality $4{\nu}^{3}=1$ is valid can be indeed obtained by considering in $d=3$ the scaling relation, 
$2{\beta}\,=\,3{\nu}\,-\,{\gamma}$ together with inequality~(\ref{eq1}):
\vspace{-0.75cm}
\begin{center}
\begin{equation} \label{eq5}
\begin{cases}
\resizebox{8.5cm}{!} {
\begin{tikzpicture}[auto] 
\node (S1) {$2{\beta}\,=\,3{\nu}\,-\,{\gamma}$};
\node (U1) [right= 2.0cm and 2.3cm of S1]
{${\gamma}\,{\leqslant}\,3{\nu}\,-\,\frac{{\nu}}{4{\nu}^{3}}\,$.};
\draw[->] (S1) to node {$2{\beta}\,{\geqslant}\,\frac{{\nu}}{4{\nu}^{3}}$} (U1); 
\end{tikzpicture} } \\
\vspace{0.1cm}
\quad \text{If}\,\,4{\nu}^{3}=1\,\,\,{\Longrightarrow}\quad {\gamma}\,{\leqslant}\,{2{\nu}} \,\,\, \text{and}\,\,\, 2{\beta}\,{\geqslant}\,{\nu}  \, .
\end{cases}
\end{equation}
\end{center}
That is, we see that inequalities ${\beta}\,{\geqslant}\,{1}/({8{\nu}^{2}})$ and  ${\gamma}\,{\leqslant}\,{2{\nu}}$ can be effectively conjugated through linear power of ${\nu}\,$. 
The latter suggests that the interpretation given by eq.~(\ref{eq5}) shows us that if there's an explicit formula for ${\nu}$, which holds for $d=3$, then inequality~(\ref{eq1}) translates the relation $2{\beta}\,=\,3{\nu}\,-\,{\gamma}$ into the following two bounds: the upper bound for ${\gamma}$ presented above as the Li-Sokal inequality and the lower bound for ${\beta}\,$ that follows from eq.~(\ref{eq2}). 

The truth of equality~(\ref{eq4}b) can be confirmed by applying the converse argumentation: To do this, let's take an the relation ${\gamma}\,=\,{\nu}\,(2\,-\,\eta)$ to transform it further.According to the proposition of Zorn's lemma~\cite{zorn} ${\gamma}$ has an upper bound ${\gamma}\,\leqslant,{2\nu}\,$. Since  $4{\nu}^{3}=1$ then
\begin{equation} \label{eq6a}
{\gamma}\,\leqslant\,\frac{1}{2\nu^{2}}
\,=\, \frac{2}{(2\nu)^{2}}\,\,\,
\overrightarrow{\,as\,\, in\,\, (\ref{eq3})} \, \,\,
\frac{2}{(2\nu+\eta^*\nu)^{2}}
\,=\, \frac{2}{(4\nu^{2}+4\nu^{2}\eta^*)}
\,=\, \frac{4{\nu}^{3}}{(2\nu^{2}+2\nu^{2}\eta^*)}
\,=\,{\nu}\,(2\,-\,{2}\eta^*).
%
\end{equation}
Hence, ${\gamma}\,\leq\,{\nu}\,(2-{2}\eta_k)$. Since ${\displaystyle \eta}$ is a limit point of 
a set of ${\displaystyle \eta_k}$ points, one suggest that to come to the mathematical equality   ${\displaystyle {\gamma}}$ must contain a point ${\displaystyle \eta}$ (as $2\eta_k\rightarrow\eta$) in the scaling relation. 

Another interesting feature of the equality $4{\nu}^{3}=1$ is that it becomes more convenient to use the resulting $\epsilon$-expansion just for ${\nu}^{-3}\,$. Due to the presence of integer value of ${\nu}^{-3}$ one expects that the faster convergent sequence of corrections terms in its $\epsilon$-representation might give a larger number of decimals. This fact was noticed in~\cite{Kompaniets2017}.   

\begin{table}[!htb]
    \caption{\label{tab:table2} Summary of estimates of the correlation length exponent ${\nu}$ and the scaling dimension ${\Delta}_{\epsilon}$ for the 3d Ising model. Third column indicates a theoretical method by which prediction has been made (CB -- conformal bootstrap, RG -- renormalization group, NRG -- non-perturbative approximation scheme to RG, BCM -- Blume-Capel model, MC -- Monte-Carlo.)}
    \centering\Large
    \resizebox{.75\columnwidth}{!}{%
    \begin{ruledtabular}
    \begin{tabularx}{0.5\linewidth}{llcl}
    ${\nu}$ & ${\Delta}_{\epsilon}$ & Method & Ref. \\[2pt] \hline \\[-10pt] 
${\sqrt[3]{2}}/{2}$ & $3-2^{2/3}$ &  & this work  \\
$(0.62996...)$ & $(1.4125989...)$ &  & \& \cite{Pishtshev2007} \\[2pt] \hline
$0.629971(4)$ & & CB & \cite{ElShowk2014} \\
           & $1.41264(6)$ & CB & \cite{SimmonsDuffin2015} \\
           & $1.412625(10)$ & CB & \cite{Kos2016} \\
$0.6292(5)   $ & & RG & \cite{Kompaniets2017}\\
$0.62977(22) $ & & RG & \cite{Shalaby2021}\\$0.63007(10) $ & & NRG & \cite{Balog2019}\\
$0.62998(5)  $ & & BCM & \cite{Hasenbusch2021}\\[2pt]
    \end{tabularx}
    \end{ruledtabular}
 }   
\end{table}
For the $d=3$ model the exponent ${\nu}$ has been estimated by different techniques with results summarized in Table~\ref{tab:table2}. 
As we can see from the comparison up to five orders of magnitude, the value given in the present work ${\nu}={\sqrt[3]{2}}/{2}=0.6299605...$ reproduces the same result as ones mathematically accessible in best schemes of the conformal bootstrap and the renormalization group.

The manipulations with cubic root of two allow one to numerically evaluate the explicit value of the anomalous dimension ${\gamma}_{\epsilon}={\Delta}_{\epsilon}-d+2$ for $d=3$ in terms of an infinite number series 
\begin{equation} \label{eq6}
{\gamma}_{\epsilon} \,=\, 
\frac{1}{3} \,+\, \frac{1}{18} \,+\, \frac{5}{324} \,+\,
\frac{5}{972} \,+\, \frac{11}{5832} \,+\, \frac{77}{104976} \,+\,
\frac{187}{629856} \,+\, \frac{935}{7558272} \,+\, \frac{21505}{408146688}
\,+\, ... \, . 
\end{equation}
The first several terms of expansion~(\ref{eq6}) being specifically rearranged change this series to the modified form affording to perform direct comparison with the for the high-precision numerical result of conformal bootstrap~\cite{Cappelli2019}: 
\begin{equation} \label{eq7}
{\gamma}_{\epsilon} =
\begin{cases}
0.3333333 + 0.0915638 - 0.0205761 + 0.0073271 + 
0.00090989 &  {\text{(from~Eq.~(\ref{eq6}))}} .\\
0.3360000 + 0.0914812 - 0.0229152 + 0.0072987 + 
0.00089005 &  {\text{(Eq.~(2.8) of Ref.~\cite{Cappelli2019})}} .
\end{cases}
\end{equation}
The first line of Eq.~(\ref{eq7}) is the modification of~(\ref{eq6}) performed in terms of regrouping and partial summations of the several series members. The second line of~(\ref{eq7}) presents the fitting polynomial-type expression~(2.8) of Ref.~\cite{Cappelli2019} written for the case $d=3\,$. Both these expressions can be matched with each other. The key result of the comparison is that as within reasonable numerical uncertainty, showing some minimal deviations, the discrepancies are insignificant, what allows us to recognize that the expressions correspond to the same value. A more sophisticated verification of the integrity of our result involving fractal dimensions comparison will determine whether the exact values of Table~\ref{tab:table1} can correspond to the model of Ref.~\cite{Cappelli2019} in which the simulation of fractional dimensions is best matched with the numerical calculations. 
In order to conduct the further comparison, we conjectured that the exact values shown in Table~\ref{tab:table1} can be merged within the following closed form:
\begin{equation} \label{eq8}
\displaystyle {\nu} \,=\, \frac{1}{2}\,\cdot\, {2}^{\,\mfrac{{4-d}}{d} }
 \,=\, \left (\frac{1}{4} \right )^{\,\mfrac{{d-2}}{d} } \, .
\end{equation}
By using then the relation ${\gamma}_{\epsilon}=2-1/{\nu}$ one can check its validity for the range of dimensions $2<d<4$. The results of comparison composed into Table~\ref{tab:table3} show a high level of the accuracy of~(\ref{eq8}).
\begin{table}[!htb]
\caption{\label{tab:table3} Comparison of ${\gamma}_{\epsilon}$ calculated by using formula~(\ref{eq8}) for fractional values of $d$ selected in the range of dimensions \thickmuskip=2.5mu $2<d<4$ with: (i) the polynomial fit of bootstrap data suggested in Ref.~\cite{Cappelli2019}, and (ii) the data of Refs.~\cite{Holovatch1993}~and~\cite{Holovatch1992} obtained within the framework of renormalization group approach. }
\centering
\resizebox{\columnwidth}{!}{%
    \begin{ruledtabular}
\begin{tabular}{l|llllllll|ccc} 
${\gamma}_{\epsilon}=2-1/{\nu}$ & \thickmuskip=2.5mu $d=2.2$ & \thickmuskip=2.5mu \thickmuskip=2.5mu $d=2.4$ & \thickmuskip=2.5mu $d=2.6$ & \thickmuskip=2.5mu $d=2.8$ & \thickmuskip=2.5mu $d=3.2$ & \thickmuskip=2.5mu $d=3.4$ & \thickmuskip=2.5mu $d=3.6$ & \thickmuskip=2.5mu $d=3.8$ & \thickmuskip=2.5mu $d=2$ & \thickmuskip=2.5mu $d=3$ & \thickmuskip=2.5mu $d=4.0$ \\ \hline
${\nu}$ from Eq.~(\ref{eq8})  & $0.8657$ & $0.7401$ & $0.6230$ & $0.5140$ & $0.3182$ & $0.2303$ & $0.1482$ & $0.0716$ & $1$ & $0.4126$ & $0$ \\ \hline
Ref.~\cite{Cappelli2019} & $0.8610$ & $0.7351$ & $0.6196$ & $0.5127$ & $0.3189$ & $0.2306$ & $0.1478$ & $0.0707$ & $0.9999$ & $0.4127$ & $0$ \\ [-4pt]
(Eq.~(2.8)) & & & & & & & & & & & \\ \hline  
Ref.~\cite{Holovatch1993} & $0.866$ & $0.741$ & $0.624$ & $0.514$ & $0.316$ & $0.227$ & $0.148$ & $0.069$ & $1$ & $0.4127$ & $-$ \\ \hline
Ref.~\cite{Holovatch1992} & $0.856$ & $0.736$ & $0.624$ & $0.518$ & $0.322$ & $0.233$ & $0.148$ & $0.069$ & $0.981$ & $0.418$ & $-$
\end{tabular}%
    \end{ruledtabular}
}%
\end{table}
One can emphasize a very good consistency between the values determined from
formula~(\ref{eq8}) and those of simulated fit data obtained in~\cite{Cappelli2019} by means of bootstrap method. 
The last row of Table~\ref{tab:table3} which additionally compares our quotes for ${\gamma}_{\epsilon}$ with predictions provided for the ${\nu}$ values by renormalization group computations in three-loop approximation with account of the exact data~\cite{Holovatch1993} supports this result as well. That is, this means that expression~(\ref{eq8}) is adequate because it is clearly seen how accurately in the whole range of $d$ between $2$ and $4$ the suggested form of~(\ref{eq8}) reproduces numerical estimates similar to those presented in~\cite{Holovatch1993}.

It is also interesting to characterize the functional behavior of~(\ref{eq8}) by analyzing its linear asymptotics with respect to $d\,$, i.e. when $d$ is lowered below four. In fact, the expression~(\ref{eq8}) consists of a product of two factors -- the first one represents the mean-field result, $1/2$, and the second factor introduces the correction that in an exponential manner depends on dimension of the system. Near integer dimensions $d$$\,=\,$$2$, $3$ and $4$ (\ref{eq8}) can be decomposed in terms of linear deviation from the integer dimension:
\begin{equation} \label{eq9}
{\nu} \,=\,
\begin{cases}
\begin{array}{lcl}
\displaystyle 
 \frac{1}{2} \,+\, \frac{1}{4}\,\cdot\,\frac{ln\,2}{2}\,\cdot \,(4-d) \,+\, O((4-d)^2) & & \text{if } d\,{\rightarrow}\,4 \\[10pt]
 \frac{\sqrt[3]{2}}{2} \,+\, \frac{4\,{\sqrt[3]{2}}}{9}\,\cdot\,\frac{ln\,2}{2}\,\cdot \,(3-d)\,+\, O((3-d)^2) & & \text{if } d\,{\rightarrow}\,3 \\[5pt]
 1 \,+\, 2\,\cdot\,\frac{ln\,2}{2}\,\cdot\, (2-d) \,+\, O((2-d)^2) & & \text{if } d\,{\rightarrow}\,2 
\end{array}
\end{cases}
\end{equation}
It is seen that when $d$ becomes smaller than four, the critical fluctuations become considerably stronger so that the weak-coupling regime of fluctuations exists only in a fairly restricted region near the \thickmuskip=2.5mu $d=4$ point.

Further outline how one can recognize additional information about a way of geometrical interpretation of $\nu$. For $d=3$ of the configurational space an explicit value of $\nu$ can be seen as particular closed form given by the following expression
\begin{equation} \label{eq10} 
\displaystyle {\nu} \,=\, {\sqrt[3]{2}}/{2}  \,=\, {exp}\left(\,{b_{4}(2)}/{6}\right)\, ,
\end{equation}
where ${\displaystyle b_{4}(2)=-\,4\,ln\,2}$ is the Madelung-type constant. According to~\cite{Borwein1987} this constant can be represented as a four-dimensional form of rectangular summations over the regular 4D cubic lattice: 
\begin{align} \label{eq11}
\displaystyle b_{4}(2) & = \sideset{}{'}\sum\limits_{i,\,j,\,k,\,l\,=\,-\,{\infty}}^{{\infty}}\,\frac{(-1)^{i+j+k+l}} {{i^2}+{j^2}+{k^2}+{l^2}} \, .
\end{align}
Lattice sum standing in the right part of~(\ref{eq10}) can be rewritten in the mixed summation-factorization form
\begin{align} \label{eq11}
\displaystyle {exp}\left(\,{b_{4}(2)}/{6}\right) & =\,  
\displaystyle 
{\nu} \,=\, 
1 \,-\, \sideset{}{'}\sum\limits_{i,\,j,\,k,\,l\,=\,-\,{\infty}}^{{\infty}}\,\, \sideset{}{'}\prod_{i,\,j,\,k,\,l\,=\,-\,{\infty}}^{{\infty}}\, \left[\, 1 \,-\, exp\,\left(\frac{1}{6}\, \frac{(-1)^{i+j+k+l}} {{i^2}+{j^2}+{k^2}+{l^2}} \right) \, \right] \, .
\end{align}
Geometrical interpretation of~(\ref{eq11}) as a physical model is simple -- it can be considered as an object assigning of a collection (convex subset) of distinct lattice points embedded orderly in 4-dimensional Euclidean space. From a homotopy equivalence $\mathbb{R}^4 \setminus \{0\} \simeq S^3$ follows that the set of points of~(\ref{eq11}) all are equidistant from a fixed central point.

In this additional part we will derive precise  evaluations of the critical exponents $\beta$ and $\gamma$ based on some Keith Conrad's expository notes on Galois theory~\cite{conrad1,conrad2}. 
First mathematical definitions. \\
Let us consider  the splitting field of $X^3 - 2$ \textbf{over $\mathbb{Q}$}.
A number belongs to $\mathbb{Q}$ if it can be expressed as a quotient or fraction of two integers:
\[
\mathbb{Q} = \left\{ \frac{a}{b} \;\middle|\; a, b \in \mathbb{Z}, \, b \neq 0 \right\} ,
\]
where $\mathbb{Z}$ denotes the set of integers ($\dots, -2, -1, 0, 1, 2, \dots$). The letter \textbf{Q} originates from the word \emph{quotient} (Giuseppe Peano, 1895). The role in Galois theory is that when ``the splitting field of $X^3 - 2$ \textbf{over $\mathbb{Q}$}'' is introduced $\mathbb{Q}$ serves as the ground field containing the coefficients of $X^3 - 2$ (namely $1$ and $-2$) and all polynomial operations, extension degrees, and automorphisms are evaluated relative to the rational numbers. Also note that a \emph{splitting field} of a polynomial $f(X)$ over a base field $K$ is the smallest field extension $L \supset K$ over which $f(X)$ factors completely into linear factors.
According to~\cite{conrad2} for $f(X) = X^3 - 2$ over $\mathbb{Q}$, the complex roots are:
\begin{align}
   r_1 = \sqrt[3]{2}\,\,\, (the\,\, real \,\,cube root)\,\,\,\,\,\, r_2 = \omega\, \sqrt[3]{2} \,\,\,\,\,\,
   r_3 = \omega^2\, \sqrt[3]{2}\,,
\end{align}
where $\omega = e^{2\pi i / 3} = \frac{-1 + i\sqrt{3}}{2}$ is a primitive non-trivial cube root of unity.
Adjoining only the real root gives the field extension $\mathbb{Q}(\sqrt[3]{2})$. Since $\mathbb{Q}(\sqrt[3]{2}) \subset \mathbb{R}$, it cannot contain the non-real roots $r_2$ and $r_3$.
To obtain the full splitting field $L$, we must adjoin both $\sqrt[3]{2}$ and $\omega$:
\[
L = \mathbb{Q}(\sqrt[3]{2},\, \omega)\,.
\]
In the context of intermediate subfields, by the Fundamental theorem of Galois theory, subfields of $L$ correspond bijectively to subgroups of $S_3$: 
$S$$_3$ contains 3 subgroups of order 2 (generated by transpositions), corresponding to the 3 cubic subfields:
    \[
    \mathbb{Q}(\sqrt[3]{2}), \quad \mathbb{Q}(\omega\sqrt[3]{2}), \quad \mathbb{Q}(\omega^2\sqrt[3]{2})
   . \]
Since number field $K~=~\mathbb{Q}(\sqrt[3]{2})$, the ring~$\mathcal{O}_K$ has unique factorization, where $\mathcal{O}_K~=~\mathbb{Z}[\sqrt[3]{2}]$~\cite{conrad2}.  
If
\[
\phi = a + b \sqrt[3]{2} + c \sqrt[3]{4}
\]
represents an algebraic integer, with $a, b, c \in \mathbb{Q}$, from computations $\Tr_{K/\mathbb{Q}}(\alpha)$, $\Tr_{K/\mathbb{Q}}(\alpha \sqrt[3]{2})$, 
and $\Tr_{K/\mathbb{Q}}(\alpha \sqrt[3]{4})$, one can see that~\cite{conrad2}
    \[
     3a, 6b, 6c \in \mathbb{Z}
   . \] \vspace{-1cm}

On the base of the above results this condition is very useful because allows us to pick out suitble coefficients in the general expression for $\phi$. Projection of $\phi$ from the point~$\mathbf{0}$ to the constrained numerical line selects scenarios maped to the denominator $6$ and without leaving it. As a result of searching coefficients with the multiplication with the denominator~$6$ we composed the following expressions for critical exponents
\begin{align*} 
& \hspace{0cm} 2\,\nu\,=\,{\sqrt[3]{2}} \quad\leftarrow \quad a=0,\,\, b=\frac{6}{6},\,\, c=0\,;\\
\hfill
& 2\,\gamma\,=\,\frac{4}{3}\,\sqrt[3]{2} \,+\, \frac{1}{2}\,{\sqrt[3]{4}}{} \quad\leftarrow \quad a=0,\,\, b=\frac{8}{6},\,\, c=\frac{3}{6}\,; \refstepcounter{equation}\tag{\theequation}\label{eq:exact}\\
\hfill
& \hspace{0.00cm} 4\, \beta\,=\,\frac{5}{3}\,\sqrt[3]{2} \,-\, \frac{1}{2}\, {\sqrt[3]{4}}{} \quad\leftarrow \quad a=0,\,\, b=\frac{10}{6},\,\, c=\,-\,\frac{3}{6}\,.
\end{align*}
\vspace{0cm}
\begin{table}[!hbt]
    \caption{\label{tab:table5} \textbf D {$=\,3\,$}: Equivalence results and the respective exact values of the correlation length exponent ${\nu}$.} 
    \centering \Large 
\resizebox{0.75\columnwidth}{!}{%
    \begin{ruledtabular}
    \begin{tabularx}{0.5\linewidth}{lcl} 
   Eq. & \hspace{-2.0cm} "Exact" value & \hspace{-0cm} Numerical value  \\[10pt] 
   $\gamma \,=\, \frac{2}{3}\sqrt[3]{2} \,+\, \frac{\sqrt[3]{4}}{4}$ & \hspace{-2.0cm} $1.23679763$  & $1.23710476$ \cite{ElShowk2014}  $\label{ref:4}{}$ \hspace{0cm} ~(a) \\[10pt]
 $\beta \,=\, \frac{5}{12}\sqrt[3]{2} \,-\, \frac{1}{8}\sqrt[3]{4}$  & \hspace{-1.85cm} $0.32654197$  &{\hspace{0cm}  \label{ref:4} $0.32643249$} \cite{ElShowk2014} \hspace{0.0cm} ~(b) \\[15pt] 
    \end{tabularx}    \end{ruledtabular}%
}    
\end{table} 
Thus, for  $d\,=\,3$ there is a direct way to think of equations~(\ref{eq:exact}) mathematically constructed  in this part correspond to exact analogues of critical exponents. Table \ref{tab:table5} states these.
\vspace{2.7cm}

\bibliography{indexes_bib}

\end{document}